\documentclass[showpacs,amsmath,amssymb,twocolumn,floatfix]{revtex4}

\usepackage{graphicx}
\usepackage{amssymb}
\usepackage{amsmath}
\usepackage{pstricks}
\usepackage[utf8]{inputenc}

\newcommand{\ket}[1]{|#1\rangle}
\newcommand{\bra}[1]{\langle #1|}
\newcommand{\cumulant}[1]{\ev{#1}_c}
\newcommand{\ev}[1]{\langle #1 \rangle}

\begin{document}

\title{Steady-state superradiance with alkaline earth atoms}

\author{D. Meiser and M. J. Holland}

\affiliation{JILA and Department of Physics, The University of
  Colorado, Boulder, Colorado 80309-0440, USA}

\date{\today}


\newcommand{\commentary}[1]{ { \texttt{#1}  } }

\begin{abstract}
  Earth-alkaline-like atoms with ultra-narrow transitions open the
  door to a new regime of cavity quantum electrodynamics. That regime
  is characterized by a critical photon number that is many orders of
  magnitude smaller than what can be achieved in conventional systems.
  We show that it is possible to achieve superradiance in steady state
  with such systems.  We discuss the basic underlying mechanisms as
  well as the key experimental requirements.
\end{abstract}

\maketitle

Superradiance, first introduced by Dicke over 50 years
ago~\cite{Dicke:Superradiance}, is one of the pillars of cavity
quantum electrodynamics (CQED). Superradiance occurs due to the
constructive interference of the probability amplitudes for
spontaneous decay of several atoms. Due to its generality and
conceptual simplicity it is a paradigm system for collective
behavior. Usually superradiance is transient; atoms initially prepared
in the excited state relax to the ground state rapidly and the
collective emission terminates.  To date superradiance has not been
achieved in a continuous fashion. The goal of this letter is to show
that steady-state superradiance can be achieved with ultracold
alkaline earth atoms in high finesse cavities. Such systems are
experimentally available in the form of optical lattice clocks
\cite{ADLudlow03282008EtAl,LemkeEtAl:YbLatticeClock}.  Recently Bose
Einstein condensates of such atoms have also become available with Ca
\cite{Kraft:CaBEC} and Sr atoms
\cite{Stellmer:StrontiumBEC,Killian:StrontiumBEC}.

Atoms with a two-electron level structure possess narrow
inter-combination lines that, due to selection rules, are
dipole-forbidden, and typically have lifetimes many orders
of magnitude longer than dipole-allowed transitions. The
long lived excited state is essential for superradiance in
steady-state because it allows the buildup of population
inversion even as the population of the excited state is
drained by the collective decay. For dipole-allowed
transitions, on the other hand, superradiant decay is so
rapid that it would exhaust the supply of excited state
atoms before it could be replenished by repumping, and
consequently the superradiant emission must cease.

\begin{figure}[!h]
  \begin{center}
    \includegraphics[width=6cm,trim=0mm 0mm 0mm
    0mm,clip=true]{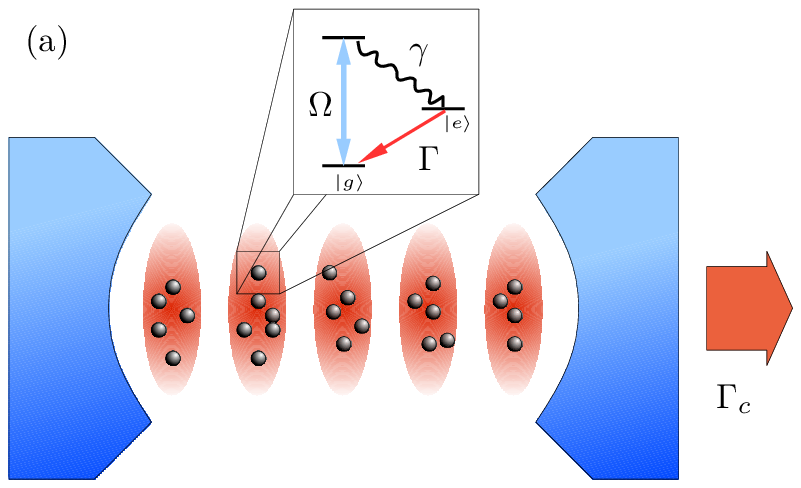}\\[0.3cm]
    \includegraphics[width=6cm]{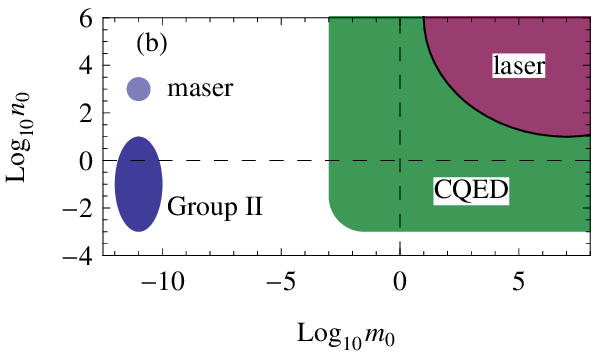}
  \end{center}
  \caption{(Color online) (a) Schematic of a coupled atom-cavity
    system leading to steady state superradiance.  (b) The parameter
    space of CQED spanned by the critical atom number $n_0$ and the
    critical photon number $m_0$.  The collective behavior of the
    coupled atom-cavity system undergoes a cross-over from stimulated
    emission dominated (laser-like) at large $m_0$ to collective
    spontaneous emission dominated (superradiance-like) for small
    $m_0$.}
  \label{fig:Schematic}
\end{figure}

Besides being of fundamental importance, steady state
superradiant systems are also interesting because of their
potential applications. The most immediate application is
the possibility to build \emph{active} optical clocks where
the light serving as a frequency standard is derived
directly from the atoms
\cite{MeiserEtAl:SrLaser,Chen:ChineseScienceBulletin}. Such
systems have the potential to improve the stability of the
best clocks by about two orders of magnitude. Another area
of application is to strongly-correlated physics. The atoms
in this system evolve into exotic many-particle states due
to their collective interaction with the light field. These
states could be of interest for quantum information purposes
as well as for the exploration and study of many-particle
phases of condensed matter.
 
The model that we consider is depicted in
Fig.~\ref{fig:Schematic}(a). $N$ two level atoms with
excited state $\ket{e}$ and ground state $\ket{g}$ decay
with rate $\Gamma_c=\mathcal{C}\Gamma$ through the mode of a
cavity, where $\mathcal{C}=g^2/(\Gamma\kappa)$ is the single
atom cooperativity parameter.  The transition from $\ket{e}$
to $\ket{g}$ is assumed to be a narrow inter-combination
line. The atomic free space spontaneous emission rate is
$\Gamma$, the single photon Rabi frequency is $g$, and the
cavity decay rate is $\kappa$. For simplicity we assume that
all atoms couple identically to the cavity mode. The rate
$\Gamma_c$ is assumed to be much larger than the rates for
non-collective decay processes and dephasing. That
approximation requires that $N\mathcal{C}\gg 1$. At the same
time the atoms are being \emph{non-collectively} repumped to
the excited state with an effective rate $w$. This could be
achieved by resonantly driving a transition to a third state
with Rabi frequency $\Omega$ that decays rapidly with rate
$\gamma$ to $\ket{e}$. 

There are two key experimental requirements for realizing
the physics discussed here. The most important requirement
is that the collective decay be dominant over all other
decay and decoherence processes, \textit{i.e.} $N \mathcal{C}
\Gamma \gg \Gamma, 1/T_2$, where $T_2$ is the spin dephasing
time. The second requirement is that one must be able to
repopulate the atomic excited state at a rate equal to the
collective decay rate, \textit{i.e.} it must be possible to
achieve $w\sim N\mathcal{C}\Gamma$ without significant
atomic losses.

The model under consideration is similar to typical many-atom CQED
systems that are studied theoretically and experimentally, except that
the atomic dipole moment is many orders of magnitude
smaller. Nevertheless, the small dipole moment can lead to profound
consequences. These can be characterized by the critical atom number
$n_0=(\kappa\Gamma)/(g^2)$ and the critical photon number
$m_0=(\gamma^2)/(g^2)$ that tell us to what degree quantum effects
are important: $n_0<1$ means that a single atom can substantially
affect the cavity field and $m_0<1$ means that the electrical field
corresponding to a single photon in the cavity can saturate the atomic
transition. For ultra-narrow inter-combination lines in
earth-alkaline-like atoms, critical photon numbers as small as
$m_0\sim 10^{-12}$ can realistically be achieved. These systems reside
in an exotic region of parameter space that has previously been
inaccessible.  For example for cavity QED systems with alkali atoms
the records are in the $m_0,\,n_0\sim 0.01-0.001$ range. It is
impossible to substantially improve upon these values as can be seen
by rewriting the critical photon number as $n_0=2\pi A/(F\sigma)$ and
$m_0=4\pi^2V_{\rm eff}/(Q\lambda_0^3)$ where $A$ is the cross section
of the cavity mode, $\sigma=3\lambda_0^2/(2\pi)$ is the resonant cross
section of the atoms, $\lambda_0$ is the wavelength of the resonant
light, and $F$ and $Q$ are the finesse of the cavity and the quality
factor of the atomic transition, respectively. Since the dimensionless
$V_{\rm eff}/(\lambda_0^3)$ must be at least unity on fundamental
grounds, and is typically orders of magnitude larger, the atomic
resonance $Q$ must be extraordinarily high to reach $m_0\sim
10^{-12}$, and the values of $Q$ that are possible for the optically
allowed dipole transitions in alkali atoms and similar systems are
insufficient. Incidentally, masers, which operate in the microwave
domain, have critical photon numbers similar to the narrow linewidth
atoms, but are classical due to their rather large critical atom
number, and have photon energies that are many orders of magnitude
smaller than is characteristic in the optical domain.

The evolution of the system shown in Fig.~\ref{fig:Schematic} is given
by the master equation,
\begin{align}
    \frac{d\hat \rho}{dt}&=
    -\frac{\Gamma_c}{2}\left( \hat J_+\hat J_-\hat
      \rho+\hat \rho \hat J_+\hat J_- -2 \hat J_- \hat \rho \hat
    J_+ \right) \label{masterEqn} \\
  &\quad -\frac{w}{2}\sum_{j=1}^N\left(\hat \sigma_-^{(j)}\hat
    \sigma_+^{(j)} \hat \rho + \hat \rho\hat
    \sigma_-^{(j)}\hat \sigma_+^{(j)}-2 \hat \sigma_+^{(j)}
    \hat \rho  \hat \sigma_k-^{(j)}\right).\notag
\end{align}
Here, $\hat \sigma_-^{(j)}=(\hat \sigma_+^{(j)})^\dagger =
\ket{g}\bra{e}$ is the lowering operator for atom $j$ and $\hat
J_-=(\hat J_+)^\dagger=\sum_{j=1}^N\hat \sigma_-^{(j)}$ is the
collective decay operator brought about by the coupling of the atoms
to the rapidly decaying cavity field. In deriving this master equation
we have assumed that the cavity field decays so rapidly that the mean
photon number is much smaller than unity, and thus the field has been
adiabatically eliminated. This approximation is valid for atoms with
an extremely weak dipole moment such as those that we are interested
in. The collective decay part occurs in the superradiance master
equation first introduced by Bonifacio \textit{et. al.}
\cite{Bonifacio:QuantumStatTheorySR1}; see also
\cite{Carmichael:StochasticInitiation}. The repumping can be thought
of as spontaneous ``absorption'' from the ground state to the excited
state.

We unravel the master equation into Monte Carlo wavefunction
trajectories~\cite{plenio1998qja}, and from the trajectories
$\ket{\psi(t)}$ we extract expectation values $\langle \hat
O(t)\rangle = \langle \psi(t)|\hat O\ket{\psi(t)}$ of system
observables $\hat O$.  Steady state expectation values $\langle \hat
O\rangle_{\rm SS}$ are then obtained by calculating time averages,
$\langle \hat O\rangle_{\rm SS}=\frac{1}{T}\int_{t_1}^{t_1+T}{\rm
  d}t\,\langle \hat O(t)\rangle\;,$ where $t_1$ is chosen large enough
to allow the system to settle to steady state, and $T$ is chosen long
enough for statistical errors to be controlled. We have empirically
checked that the steady state does not depend on the choice of initial
conditions.

\begin{figure}
  \begin{center}
    \includegraphics[trim=0mm 1mm 0mm 0mm,clip=true]{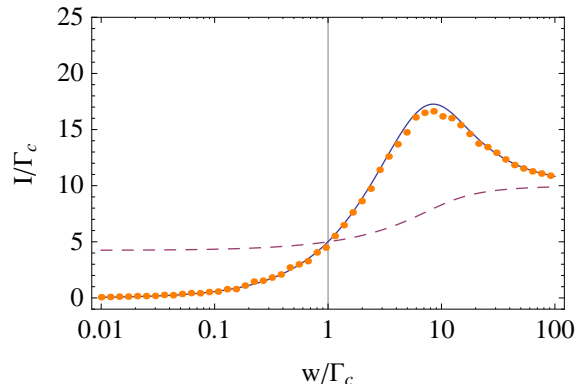}
  \end{center}
  \caption{ (Color online) Emission rate as a function of repump rate
    for 10 atoms. The orange dots are the expectation values extracted
    from the Monte-Carlo wave function simulations.  The blue line
    shows the semi-classical result obtained from
    Eqs.(\ref{eqn:sigmaz}-\ref{eqn:sigmaz1sigmaz2}), and the purple
    dashed line is the emission rate for uncorrelated particles,
    $N_e\Gamma_c$.  }
  \label{fig:eRate}
\end{figure}

In Fig.~\ref{fig:eRate} we show that Eq.~(\ref{masterEqn})
leads to sustained superradiance by investigating the mean photon
emission rate
\begin{equation}
  I=\Gamma_c\langle \hat J_+\hat J_-\rangle_{\rm SS}\,,
\end{equation}
together with the emission rate $N_e\Gamma_c$ that one would expect
for uncorrelated atoms, with $N_e$ the population of the excited
state. Three qualitatively different regimes can be distinguished:
strong pumping with $w>N\Gamma_c$, intermediate pumping with
$\Gamma_c<w<N\Gamma_c$, and weak pumping $w<\Gamma_c$. For very strong
pumping, $w\gg N\Gamma_c$ the emission rate approaches the maximum
possible emission rate for uncorrelated atoms, $N \Gamma_c$. As the
pump rate decreases, the emission rate \emph{increases} beyond $N
\Gamma_c$. This implies the presence of correlations between different
atoms, resulting in a collective enhancement of the emission rate. The
behavior of the system in the weak pumping regime, $w<\Gamma_c$ is
surprising and counter-intuitive in two ways. First, nearly half the
atoms remain in the excited state even as $w/\Gamma_c\to 0$. Second,
the emission rate is greatly suppressed below the value for
uncorrelated atoms, \textit{i.e.\/} the atoms are
subradiant. Subradiance, like superradiance, implies the presence of
correlations between different atoms.

\begin{figure}
  \begin{center}
    \includegraphics[width=8cm,trim=0mm 7mm 0mm
    6mm,clip=true]{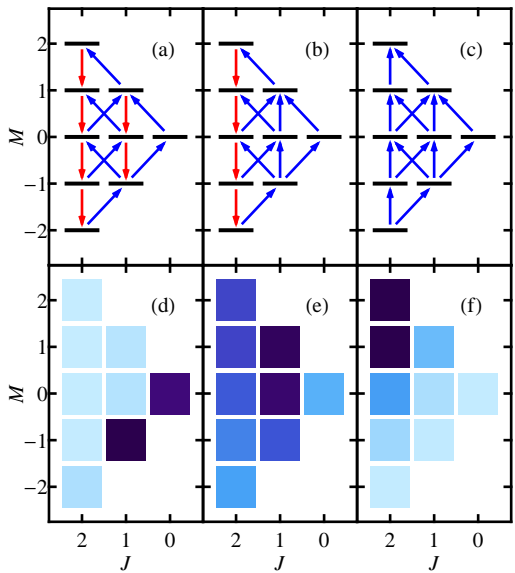}
  \end{center}
  \caption{(Color online) Transition rates between $J$ and $M$
    eigenspaces (a-c) and populations $P_{M,J}$ of these subspaces
    (d-f) for $w=0.1\,\Gamma_c$ (a, d), $w=2.0\,\Gamma_c$ (b, e),
    $w=10.0\,\Gamma_c$ (c, f), and $N=4$ atoms. Red arrows indicate
    decay dominated transitions and blue arrows indicate repump
    dominated transitions. The transition rates are calculated by
    averaging over degenerate initial states and summing over final
    states. In (d-f) darker shades indicate larger population.}
  \label{fig:LevelScheme}
\end{figure}

The qualitative behavior of the emission rate can be understood if we
look at the dynamics of the system in the collective basis
$\ket{J,M,\xi}$ where $J$ characterizes the total angular momentum,
the magnetic quantum number $M$ characterizes the inversion, and the
multiplicity quantum number $\xi$ enumerates the degenerate manifold
at given $J$ and $M$. Figures \ref{fig:LevelScheme}(a-c) show the
subspaces corresponding to well defined $J$ and $M$ along with the net
transition rates, for $N=4$ atoms. Panels (d-f) show the steady state
probabilities, $ P_{M,J}= \langle \hat P_{M,J}\rangle_{\rm SS}$ where
$\hat P_{M,J}= \sum_\xi |J,M,\xi\rangle\langle J,M,\xi|$, for the
system to be in the various subspaces. The right hand column (panel c
and f) of Fig.~\ref{fig:LevelScheme} shows the strong pumping limit
where repumping dominates the collective decay and consequently a
large fraction of the atoms accumulate in the excited state.

In the intermediate pumping regime, $w\lesssim N\Gamma_c$, shown in
the central column (panel b and e), there is a non-trivial competition
between collective decay and repumping. It is apparent that for states
where $J \sim \mathcal{O}(N/2)$ and is therefore close to its maximum
possible value, the collective decay dominates, whereas for states
with small $J$ the non-collective repumping dominates. This leads to a
cycle in which the decay and repumping balance with an enhanced
average value for $\langle J_+J_-\rangle$, {\em i.e.\/} an enhanced
emission rate. Note that it is important that the repumping is
non-collective for superradiance to occur in steady state. If the
repumping also preserves $J$, as is the situation considered in
Refs.~\cite{Haake:SuperradiantLaser,
  HaakeEtAl:SuperradiantLaserPartialInv,
  HaakeEtAl:NoiseReductionStationarySuperradiance}, there is no
mechanism to balance the non-collective decay and decoherence
processes that tend to decrease $J$ and are unavoidable in experiment.

In the weak pumping regime, shown in the left hand column (panel a and
d), the collective emission drives the system into states with $M=-J$
that cannot decay by means of $\hat J_-$. For $M < 0$ the repumping
predominantly drives transitions $J \to J - 1$ while transitions from
$J \to J + 1$ are rare. Thus the system evolves along a dynamical
pathway with smaller and smaller $J$, eventually reaching $J\sim 1$
where it becomes trapped. This steady state has an almost equal number
of atoms in the ground and excited state, and yet has a greatly
suppressed emission rate. This is due to subradiance arising from
strong atom-atom correlations. The requirement for this subradiance to
occur is $\mathcal{C}\gg 1$, which is much more stringent than the
condition for superradiance.

The Monte Carlo simulations provide a clear picture of the dynamical
interference, and are complete in that they allow us to include a full
description of the Hilbert space evolution. Due to the exponential
scaling of the dimension of the Hilbert space of the system with the
number of particles we are however limited to relatively small numbers
of particles (of order 20). To consider mesoscopic particle numbers,
we use a semi-classical approximation that consists of keeping only
pair correlations. This approximation is validated by the excellent
agreement with the Monte-Carlo results for small particle
numbers. Mathematically, the semi-classical approximation is
implemented by expanding expectation values of the system operators
$\{\hat \sigma_z^{(j)},\hat \sigma_+^{(j)},\hat \sigma_-^{(j)}\}$ in
terms of cumulants $\langle \ldots\rangle_c$, where $\hat
\sigma_z^{(j)} = \ket{e}\bra{e} - \ket{g}\bra{g}$. We use that all
expectation values are symmetrical with respect to particle exchange,
\textit{e.g.}  $\langle \hat \sigma_+^{(i)}\hat \sigma_-^{(j)}\rangle
= \langle \hat \sigma_+^{(1)}\hat \sigma_-^{(2)}\rangle$ for all
$i\neq j$. All non-zero cumulants up to second order can be expressed
in terms of $\langle \hat \sigma_z^{(1)}\rangle_c = \langle \hat
\sigma_z^{(1)}\rangle$, $\langle \hat \sigma_+^{(1)}\hat
\sigma_-^{(2)}\rangle_c = \langle \hat \sigma_+^{(1)}\hat
\sigma_-^{(2)}\rangle$, and $\langle \hat \sigma_z^{(1)}\hat
\sigma_z^{(2)}\rangle_c = \langle \hat \sigma_z^{(1)}\hat
\sigma_z^{(2)}\rangle- \langle \hat \sigma_z^{(1)}\rangle^2 $ and
their equations of motion are
\begin{align}
  \frac{d\langle \hat
  \sigma_z^{(1)}\rangle_c}{dt}&=      \label{eqn:sigmaz}
  -(w+\Gamma_c)\langle \hat
  \sigma_z^{(1)}\rangle_c\\
 &\quad -2\Gamma_c(N-1)\langle \hat \sigma_+^{(1)}\hat
  \sigma_-^{(2)}\rangle_c\;,\nonumber
\end{align}
\begin{align}
  \frac{d\langle \hat\sigma_+^{(1)}\hat
  \sigma_-^{(2)}\rangle_c}{dt}&=
  -(w+\Gamma_c)\langle \hat
  \sigma_+^{(1)}\sigma_-^{(2)}\rangle_c\\
  &\quad+\frac{\Gamma_c}{2}\left(\langle\hat
  \sigma_z^{(1)}\hat\sigma_z^{(2)}\rangle_c+\langle \hat
  \sigma_z^{(1)}\rangle_c\right)\nonumber\\
  &\quad+\Gamma_c(N-2)\Big(\langle\hat\sigma_z^{(1)}\rangle_c\langle
  \hat\sigma_+^{(1)}\hat\sigma_-^{(2)}\rangle_c\nonumber\\
  &\qquad\quad+\langle\hat\sigma_z^{(1)}
  \hat\sigma_+^{(2)}\hat\sigma_-^{(3)}\rangle_c\Big)\;,\nonumber
\end{align}
\begin{align}
  \frac{d\langle
  \hat \sigma_z^{(1)}\hat \sigma_z^{(2)}\rangle}{dt}&=
  \label{eqn:sigmaz1sigmaz2} -2(w+\Gamma_c)\langle
  \hat \sigma_z^{(1)}\hat \sigma_z^{(2)}\rangle_c\\ 
&\quad +4\Gamma_c\Big(\langle
  \hat \sigma_+^{(1)}\hat \sigma_-^{(2)}\rangle_c(1+\langle \hat
  \sigma_z^{(1)}\rangle_c)\nonumber\\ 
&\qquad\quad-(N-2)\langle \hat
  \sigma_z^{(1)}\hat \sigma_+^{(2)}\hat
  \sigma_-^{(3)}\rangle_c\Big)\;.\nonumber
\end{align}
By dropping the small third order cumulant $\langle \hat
\sigma_z^{(1)}\hat \sigma_+^{(2)}\hat \sigma_-^{(3)}\rangle_c$ we
obtain a closed set of equations. The steady state is found by setting
the time derivatives to zero. The resulting cubic equations can be
solved exactly and these solutions are the basis of the analytical
curve in Fig.~(\ref{fig:eRate}). The expressions one obtains are
however very complicated. Simple expressions may be obtained in the
regime $\Gamma_c\ll w\sim N\Gamma_c$ in which collective emission
occurs. By introducing the rescaled operators $\hat j_z=\hat J_z/N$
and $\hat j_\pm =\hat J_\pm /N$ where $\hat J_z=1/2\sum_j\hat
\sigma_z^{(j)}$, we find that to leading order in $1/N$ the only
non-zero expectation values are $\cumulant{\hat j_z}$ and
$\cumulant{\hat j_+\hat j_-}$ and they evolve according to,
\begin{eqnarray}
  \frac{d\cumulant{\hat j_z}}{dt}&=&
  -w(\cumulant{\hat j_z}-1/2)-N\Gamma_c\cumulant{\hat j_+\hat j_-}\\
  \frac{d\cumulant{\hat j_+\hat j_-}}{dt}&=&-w\cumulant{\hat
  j_+\hat j_-}+2N\Gamma_c\cumulant{\hat j_z}\cumulant{\hat
  j_+\hat j_-}.
\end{eqnarray}
The steady state solutions are
\begin{equation}
  \langle \hat j_z\rangle_{\rm SS}=
  \begin{cases}
    \frac{1}{2}\frac{w}{N\Gamma_c}\;,&
    w<N\Gamma_c\\
    \frac{1}{2}\;,&w\geq N\Gamma_c
  \end{cases}\;,
\end{equation}
and
\begin{equation}
  \langle
  \hat j_+\hat j_-\rangle_{\rm SS}=
  \begin{cases}
    \frac{1}{2}\frac{w}{N\Gamma_c}\left(1-\frac{w}{N\Gamma_c}
    \right)\;,&w<N\Gamma_c\\
    0\;&w\geq N\Gamma_c
  \end{cases}.
  \label{Eqn:steadyStateEmissionRate}
\end{equation}

The fact that $\langle \hat j_+\hat j_-\rangle_{\rm SS}$ is of order
unity for $w\lesssim N\Gamma_c$ indicates the presence of strong
correlations between atoms. From
Eq.~(\ref{Eqn:steadyStateEmissionRate}) we can extract the maximum
intensity,
\begin{equation}
  I_{\rm max}=N^2\Gamma_c/8,
\end{equation}
obtained at $w=N\Gamma_c/2$. The scaling of that intensity with $N^2$
underlines the collective nature of the light emission. Remarkably,
$I_{\rm max}$ is only a factor $1/2$ smaller than the maximum possible
emission rate $\Gamma_c\langle J,0|\hat J_+\hat J_-|J,0\rangle$.

Not only is this source of light bright, however, it also has a long
coherence time---a property shared with lasers. In lasers, the
coherence time of the light field is much longer than the
cavity-ring-down time because of the fact that the field is
macroscopically occupied \cite{Schawlow:InfraredAndOpticalMasers}. To
find the coherence time of the atoms in the superradiant system
considered in this paper, we study the two-time correlation function
of the atomic dipole $\langle \hat \sigma_+^{(1)}(t+\tau) \hat
\sigma_-^{(2)}(t) \rangle$. Using the quantum regression theorem we
find the equations of motion
\begin{align}
  \frac{d \langle \hat
  \sigma_+^{(1)}(\tau)\hat\sigma_-^{(2)}(0)\rangle}{d\tau}=&
 -\frac{1}{2}\left(w+\Gamma_c-(N-2)\langle
  \hat \sigma_z^{(1)}\rangle_{\rm SS}\right)\nonumber\\
&\quad\mbox{}\times\langle\hat
  \sigma_+^{(1)}(\tau)\hat \sigma_-^{(2)}(0)\rangle\;,
\end{align}
where we have factorized $\langle \hat \sigma_+^{(1)}(\tau)
\hat\sigma_z^{(2)}(\tau) \hat\sigma_-^{(3)}(0) \rangle \approx \langle
\hat \sigma_+^{(1)} \rangle_{\rm SS} \langle \hat\sigma_z^{(2)}(\tau)
\hat\sigma_-^{(3)}(0) \rangle $ and we have assumed that the system is
in steady state.  Inserting $\langle \hat \sigma_z^{(1)}\rangle_{\rm
  SS}=2\langle \hat j_z\rangle_{\rm SS}$ from above we find that the
spins of different atoms remain coherent with each other for $t_{\rm
  coh}=N/(N\Gamma_c+2w)$.  In the superradiant regime where $w\sim
N\Gamma_c$ the coherence time is thus indeed a factor $\sim N$ larger
than that for just two atoms. The phase of the atomic dipole diffuses
slowly which is analogous to the slow phase diffusion of the field of
a laser. This result is significant especially with an eye to possible
applications as an ultrastable local oscillator or frequency
reference.

In summary, we have shown that steady state superradiance can be
achieved with atoms with an ultra-narrow transition in a
cavity. Alkaline-earth-like atoms are prime candidates for realizing
such systems and open up a new regime of CQED characterized by
an extremely small critical photon number.  In future work we will
study the noise properties of the emitted light and the correlated
atomic state. It will be intruiging to consider the crossover from
steady-state superradiance to lasing that occurs as the system goes
from the bad cavity limit to the good cavity limit.

We thank J. K. Thompson, H. Uys, P. Zoller, and J. Cooper
for helpful discussions. This work has been funded in part
by NSF and DOE. D. M. gratefully acknowledges support from
Deutsche Forschungsgemeinschaft.

\bibliographystyle{apsrev}
\bibliography{mybibliographyEtAl}

\end{document}